\documentclass[fleqn,usenatbib]{mnras}

\usepackage{newtxtext,newtxmath}
\usepackage[T1]{fontenc}

\DeclareRobustCommand{\VAN}[3]{#2}
\let\VANthebibliography\thebibliography
\def\thebibliography{\DeclareRobustCommand{\VAN}[3]{##3}\VANthebibliography}

\usepackage{graphicx}
\usepackage{amsmath}

\newcommand{\bfr}{{\boldsymbol{r}}}
\newcommand{\bfv}{{\boldsymbol{v}}}
\newcommand{\txa}{{\text{a}}}
\newcommand{\txA}{{\text{A}}}
\newcommand{\txB}{{\text{B}}}

\newcommand{\txd}{{\text{d}}}

\newcommand{\txt}{{\text{t}}}
\newcommand{\txT}{{\text{T}}}
\newcommand{\calE}{{\cal{E}}}
\newcommand{\calET}{{\cal{E}}_{\text{T}}}

\newcommand{\rT}{r_{\text{T}}}

\DeclareMathOperator{\floor}{floor}

\title[Uniform density sphere]{Self-consistent dynamical models with a finite extent  -- I.~The uniform density sphere}

\author[M. Baes]{Maarten Baes
\\
Sterrenkundig Observatorium, Universiteit Gent, Krijgslaan 281 S9, 9000 Gent, Belgium
}

\date{Accepted 2022 March 9. Received 2022 March 8; in original form 2022 February 24}

\pubyear{2022}

\begin{document}
\label{firstpage}
\pagerange{\pageref{firstpage}--\pageref{lastpage}}
\maketitle

\begin{abstract}
The standard method to generate dynamical models with a finite extent is to apply a truncation in binding energy to the distribution function. This approach has the disadvantages that one cannot choose the density to start with, that the important dynamical quantities cannot be calculated analytically, and that a fraction of the possible bound orbits are excluded a priori. We explore another route and start from a truncation in radius rather than a truncation in binding energy. We focus on the simplest truncated density profile, the uniform density sphere. We explore the most common inversion techniques to generate distribution functions for the uniform density sphere, corresponding to a large range of possible anisotropy profiles. We find that the uniform density sphere cannot be supported by the standard isotropic, constant anisotropy or Osipkov-Merritt models, as all these models are characterised by negative distribution functions. We generalise the Cuddeford inversion method to models with a tangential anisotropy and present a one-parameter family of dynamical models for the uniform density sphere. Each member of this family is characterised by an anisotropy profile that smoothly decreases from an arbitrary value $\beta_0\leqslant0$ at the centre to completely tangential at the outer radius. All models have a positive distribution function over the entire phase space, and a nonzero occupancy of all possible bound orbits. This shows that one can generate nontrivial self-consistent dynamical models based on preset density profile with a finite extent. 
\end{abstract}


\section{Introduction}

Simple analytical dynamical models are a useful and important tool in the study of star clusters, galaxies, galaxy clusters and dark matter haloes. They serve as an idealised environment where different physical processes can be investigated, or new modelling or data analysis techniques can be explored. Furthermore, they can be used as starting point for the creation of more complex models or full numerical simulations. For a full overview of the field, we refer to \citet{2008gady.book.....B} and \citet{2021isd..book.....C}.

In the quest for analytical dynamical models, one usually has to make a choice between simplicity and realism. A popular starting point is the assumption of spherical symmetry: a large number of dynamical models have been generated with a spherical density profile $\rho(r)$ or gravitational potential $\Psi(r)$ as starting point \citep[e.g.,][]{1959AnAp...22..126H, 1983MNRAS.202..995J, 1987MNRAS.224...13D, 1990ApJ...356..359H, 1991A&A...249...99C, 1993MNRAS.265..250D, 1994AJ....107..634T, 1996MNRAS.278..488Z, 2001MNRAS.321..155L, 2004MNRAS.351...18B, 2005MNRAS.360..492E, 2005MNRAS.358.1325C, 2020A&A...634A.109B, 2021MNRAS.503.2955B}. All of these commonly used analytical dynamical models listed above have an infinite extent, meaning that their density is nonzero over the entire 3D space. It is useful, however, to also study the dynamical properties of models with a finite extent.

The standard method to generate dynamical models with a finite extent is to take an analytical model with an infinite extent and to apply a truncation in binding energy to the distribution function (DF). By only allowing orbits with binding energies larger than a given value $\calE_{\text{T}}$, the model will automatically have a finite radial extent, with maximum or truncation radius $r_{\text{T}}$ defined by $\Psi(r_{\text{T}}) = \calE_{\text{T}}$. Probably the most famous models of this kind are the King models \citep{1966AJ.....71...64K}, which can be considered as energy-truncated versions of the isothermal sphere. This energy truncation approach can in principle be applied to any analytical distribution function, as binding energy is always one of the isolating integrals of motion. Other examples of energy truncation have been presented, either as stand-alone models or as building blocks in multi-component dynamical systems \citep[e.g.,][]{1963MNRAS.125..127M, 1970AJ.....75..674P, 1975AJ.....80..175W, 1998MNRAS.298..677D, 2004MNRAS.349..440D}. 

One disadvantage of this energy truncation method is that it generally does not generate analytical models with a preset density distribution. Even if one starts from a fully analytical model in which the most important dynamical quantities are known exactly, this does not longer apply to the truncated version. Indeed, by truncating the DF in binding energy, also the mass density changes, and as a result also the gravitational potential. The bottomline is that we only have a closed expression for the DF as a function of the integrals of motion, but as the potential is not known, we do not have explicit expressions for any of the important dynamical properties. In the best case, we can derive an explicit expression for the augmented density by integrating the DF over velocity space, after which the potential can be found numerically by solving Poisson's equation. In short, while this approach has a simple onset, typically one cannot choose the density of the model a priori. 

Another, possibly more fundamental issue of this energy truncation method has been brought up by \citet{1988ApJ...325..566K}. He argues that a truncation in binding energy imposes severe and unphysical limitations on the system described. The most important constraint relates to the type of orbits represented in the system. It seems logical for a spherical dynamical system to contain, at any radius, orbits ranging from radial to circular, provided that they are gravitationally bound to the system. In other words, it feels unnatural to exclude possible orbits from the system a priori unless they are not gravitationally bound. This argument, however, breaks down when applying an energy truncation: an energy truncation excludes all orbits with $\calE<\calE_{\text{T}}$, even when these orbits are still gravitationally bound. This applies in particular to nearly circular orbits in the outer regions. \citet{1988ApJ...325..566K} argues that dynamical systems such as globular clusters or galaxies should be truncated in radius, rather than in binding energy. 

This brings up the questions whether it is possible to generate self-consistent dynamical models truncated in radius, with a preset density profile and, if possible, with a different range of orbital structures.  In this paper we investigate in detail the dynamical structure of the simplest model with a truncation in radius: the uniform density sphere. This model is one of the classical models used in stellar dynamics \citep{1971Afz.....7..223B, 2021Ap.....64..219B, 1974SvA....17..460P, 1979PAZh....5...77O, 2008gady.book.....B} and gravitational lensing \citep{1972MNRAS.158..233C, 2002MNRAS.337.1269W, Suyama_2005}. It is well-known that the uniform density sphere cannot be supported by an isotropic velocity distribution \citep{Zeldovich72, 1979PAZh....5...77O, 2008gady.book.....B}. Quoting the latter authors: there is no stellar-dynamical analog of a self-gravitating sphere of incompressible liquid. On the other hand, it is possible to populate the uniform density sphere with purely circular orbits \citep{1971Afz.....7..223B}. In fact, the circular orbit model is the only orbital structure that is guaranteed to always yield a positive DF, and hence a physically viable dynamical model, at least in theory, for every spherical density profile \citep{1984ApJ...286...27R, 2008gady.book.....B}.

The goal of this paper is to make use of the comprehensibility of the uniform density sphere to investigate in detail the connection between a truncation in binding energy and a truncation in radius, and to search for physically viable DFs that can self-consistently generate this prototypical example of a density distribution with a finite extent. This paper is organised as follows. In Sec.~{\ref{general.sec}} we present an overview of the general properties of the uniform density sphere. In Sec~{\ref{quest.sec}} we investigate the most common approaches to construct dynamical models with a fixed density profile: we consecutively investigate the isotropic model (Sec.~{\ref{iso.sec}}), the radial and circular orbit models (Sec.~{\ref{radcirc.sec}}), models with a constant anisotropy (Sec.~{\ref{cani.sec}}), and Osikpov-Merritt models (Sec.~{\ref{OM.sec}}). In Sec.~{\ref{newfam.sec}} we present a new one-parameter family of dynamical models that self-consistently generate the uniform density sphere. Every model in this family has a positive DF over the entire phase space, and is populated by a complete mixture of all allowed orbits. We discuss and summarise our results in Sec.~{\ref{sumdisc.sec}}.

\section{General properties}
\label{general.sec}

The uniform density sphere is characterised by the almost trivial density profile
\begin{equation}
\rho(r) = \begin{cases}
\;\dfrac{3}{4\pi}\,\dfrac{M}{r_{\text{T}}^3} &\quad{\text{for }}r < r_{\text{T}},
\\[1em]
\;0
&\quad{\text{for }}r > r_{\text{T}}.
\end{cases}
\end{equation}
It contains two parameters: the total mass $M$ and the truncation radius $r_{\text{T}}$. To simplify notations, we will work in dimensionless units in the remainder of this work, and we set $G=M=\rT=1$. In these dimensionless units, we rewrite the density as
\begin{equation}
\rho(r) = 
\frac{3}{4\pi}\,\Theta(1-r),
\label{UDS-rho}
\end{equation}
with $\Theta(x)$ the Heaviside step function. 
The basic properties of the uniform density model are readily calculated: the cumulative mass profile is simply
\begin{equation}
M(r) = 
\begin{cases}
\;r^3 & \quad\text{for }r\leqslant1,
\\
\;1 & \quad\text{for }r\geqslant1.
\end{cases}
\label{UDS-M}
\end{equation}
The gravitational potential is
\begin{equation}
\Psi(r) = 
\begin{cases}
\;\frac{1}{2}\left(3-r^2\right) & \quad\text{for }r\leqslant1,
\\[0.5em]
\;r^{-1} & \quad\text{for }r\leqslant1.
\end{cases}
\label{UDS-Psi}
\end{equation}
The depth of the potential well is $\Psi_0 = \tfrac32$. The total potential energy of the uniform density sphere is well known \citep{2008gady.book.....B, 2019A&A...630A.113B}:
\begin{equation}
W_{\text{tot}} =  -2\pi\int_0^\infty \rho(r)\,\Psi(r)\,r^2\,{\text{d}}r = -\frac{3}{5}.
\label{Wtot}
\end{equation}

\begin{figure*}
\includegraphics[width=0.92\textwidth]{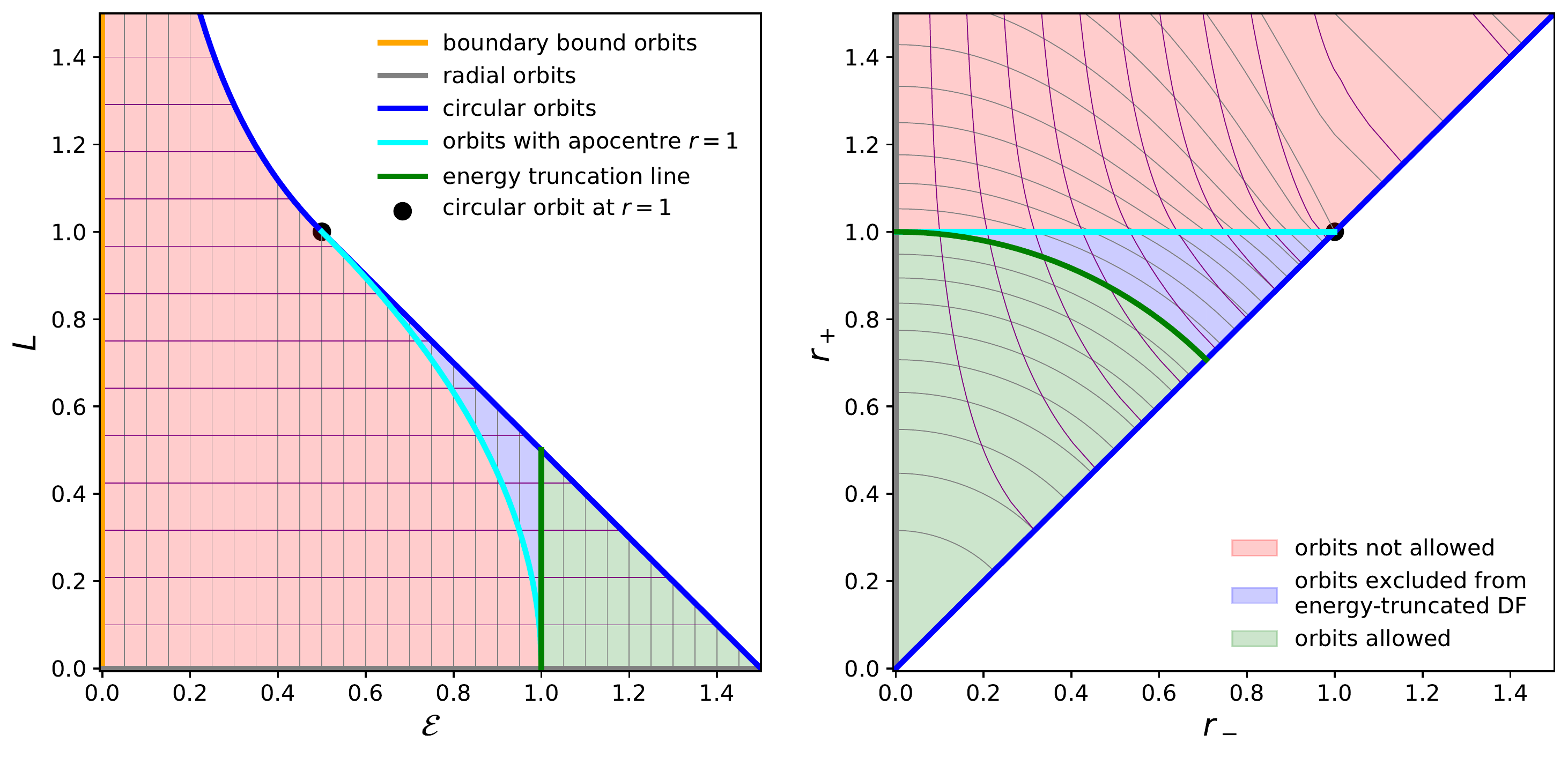}%
\hspace*{4em}%
\caption{$(\calE,L)$ parameter space (left) and turning point space (right) for the uniform density sphere. The green region corresponds to orbits with $1\leqslant\calE\leqslant\tfrac32$ and $r_+\leqslant 1$, which are present in the energy-truncated dynamical models. The blue region corresponds to orbits with $\calE<1$ and $r_+\leqslant 1$, which are allowed in the uniform density sphere but excluded from energy-truncated models. The pink region contains orbits with $r_+>1$, which are not allowed in the uniform density sphere. The thin grey and purple lines correspond to lines of constant binding energy and binding angular momentum, respectively.}
\label{paramspace.fig}
\end{figure*}

In a static spherically symmetric system, each star moves on an planar orbit characterised by the total binding energy and the total angular momentum (both per unit mass),
\begin{gather}
\calE = \Psi(r) - \tfrac12\left(v_r^2+v_{\text{t}}^2\right), \label{defE}\\
L = r\,v_{\text{t}}. \label{defL}
\end{gather}
Each gravitationally bound orbit describes a rosetta  in the orbital plane, with a pericentre $r_-$ and apocentre $r_+$. According to Bertrand's theorem \citep{wear2010classical}, the point potential and the radial harmonic oscillator potential are the only two central-force potentials in which all bound orbits are closed. The potential (\ref{UDS-Psi}) is a combination of both these cases. Bound orbits with $r_+\leqslant1$ are elliptical orbits with the centre of the system as the geometric centre of the orbit, whereas bound orbits with $r_-\geqslant1$ are elliptical orbits with the centre of the system a focal point of the ellipse. Bound orbits with $r_-<1<r_+$ have a more general rosetta shape. 

As the density of the uniform density sphere disappears beyond $r=1$, we focus on orbits with $r_+\leqslant1$. For such orbits, we can explicitly calculate the connection between the integrals of motion $(\calE,L)$ and the turning points $(r_-,r_+)$ of each orbit are found by setting $v_r=0$ in expression~(\ref{defE}), substituting Eqs.~(\ref{UDS-Psi}) and (\ref{defL}) and solving for $r$,
\begin{equation}
r_\pm^2(\calE,L) = \frac{3-2\calE}{2} \pm \sqrt{\left(\frac{3-2\calE}{2}\right)^2-L^2}.
\label{periapo}
\end{equation}
The inverse relations are
\begin{gather}
\calE =  \tfrac12\left(3-r_-^2-r_+^2\right),
\\
L = r_-\,r_+.
\end{gather}
The condition $r_+\leqslant1$ translates to
\begin{equation}
\left(\frac12\leqslant\calE\leqslant1 \;{\text{and}}\; L\leqslant\sqrt{2-2\calE}\right)
\;{\text{or}}\;
\left(1\leqslant\calE\leqslant\frac32\right). 
\end{equation}
In Fig.~{\ref{paramspace.fig}} we show the $(\calE,L)$ parameter space (left) and the turning point space (right) for the uniform density sphere. The colours of the lines and regions in the two plots correspond to each other. The boundary between gravitationally bound and unbound orbits is determined by the condition $\calE=0$ or $r_+\to\infty$. At fixed $\calE$, the angular momentum can range from $L=0$, the value for radial orbits, to 
\begin{gather}
L_{\text{c}}(\calE) = 
\begin{cases}
\;\dfrac{1}{\sqrt{2\calE}} & \quad{\text{for }}0<\calE\leqslant\tfrac12,
\\[1.4em]
\;\dfrac{3-2\calE}{2} & \quad{\text{for }}\tfrac12\leqslant\calE\leqslant\tfrac32,
\end{cases}
\label{LcE}
\end{gather}
corresponding to circular orbits. The energy truncation method would truncate the range of possible binding energies to $1\leqslant\calE\leqslant\tfrac32$, indicated as the green region in both plots. All tightly bound orbits with $1\leqslant\calE\leqslant\frac32$ have an apocentre smaller than the truncation radius. However, also the more loosely bound orbits in the blue region correspond to bound orbits with $r_+<1$. For $\frac12 \leqslant \calE \leqslant 1$, the most radial orbits are prohibited since they reach radii beyond the truncation radius. As $\calE$ decreases and reaches $\calE=\tfrac12$, only circular orbits are possible. All orbits with binding energy below this limiting value pass beyond the truncation radius and are excluded.

\section{A quest for viable dynamical models}
\label{quest.sec}

\subsection{Isotropic model}
\label{iso.sec}

In general, dynamical models are fully described by their phase-space DF $f(\bfr,\bfv)$. In spherical symmetry, the DF can be written as $f(\bfr,\bfv) = f(\calE,L)$, that is, it depends on the six phase-space coordinates only through the binding energy and the angular momentum. An important subclass are isotropic or ergodic dynamical models, in which the DF only depends on the binding energy. Such models are an evident choice in the sense that there is a relatively simple relation between density and DF. \citet{1986PhR...133..217D} argued that isotropic DFs are also the evident first option based on the maximum entropy principle. 

When the density and potential are known, the functional form of the unique ergodic DF can be calculated through the standard Eddington formula, 
\begin{equation}
f(\calE) = \frac{\sqrt2}{4\pi^2} \frac{\txd}{\txd\calE} \int_0^\calE \frac{\txd\tilde\rho}{\txd\Psi}\,\frac{\txd\Psi}{\sqrt{\calE-\Psi}},
\label{eddington}
\end{equation}
where $\tilde\rho(\Psi)$ is the augmented density, that is, the density written as a function of the potential. For the uniform density sphere, $\tilde\rho(\Psi)$ can be written as
\begin{equation}
\tilde\rho(\Psi) = \frac{3}{4\pi}\,\Theta(\Psi-1).
\label{UDS-ad}
\end{equation} 
The Heaviside step function is not differentiable in the classical sense, but we can formally write the derivative of the augmented density in terms of the Dirac delta distribution,
\begin{equation}
\frac{\txd\tilde\rho}{\txd\Psi}(\Psi) = \frac{3}{4\pi}\,\delta(\Psi-1).
\end{equation} 
Substituting the last expression in the Eddington equation~(\ref{eddington}) leads to the following formal expression for the DF,
\begin{equation}
f(\calE) 
=
\frac{3\sqrt2}{16\pi^3}
\left[ \frac{\delta(\calE-1)}{\sqrt{\calE-1}} - 
\frac{\Theta(\calE-1)}{2\,(\calE-1)^{3/2}}\right].
\label{df-uniformdensity}
\end{equation}
The formal expression (\ref{df-uniformdensity}) for the DF has a number of interesting characteristics: it vanishes for all $\calE<1$, it has an infinite peak at $\calE=1$, and it is negative for all binding energies $1<\calE\leqslant\tfrac32$. It is clear that this is not a proper probability density function in phase space, which confirms previous studies that the uniform density sphere cannot be supported by an ergodic DF. 

To understand these characteristics, it is useful to look at a dynamical model as a superposition of orbits, as in Schwarzschild orbit superposition methods \citep{1979ApJ...232..236S, 1984ApJ...286...27R, 1997ApJ...488..702R, 2021MNRAS.500.1437N}. Every orbit is uniquely characterised by its pericentre $r_-$ and apocentre $r_+$, and we can calculate the mass density and the velocity dispersion profiles corresponding to every orbit. Building up a dynamical model comes down to finding the relative mass contribution for each orbit by requiring that the cumulative contribution of all the orbits agrees with the desired density profile and velocity dispersion profiles. 

We start building up the model from the outside and gradually move inward. We obviously neglect all orbits with $r_+>1$. Of all the possible orbits with $r_+=1$, we can only expect a contribution of purely radial orbits. Indeed, all orbits with $r_+=1$ have zero radial velocity here, implying that the radial velocity dispersion is zero. If we want to maintain velocity isotropy, also the tangential velocity dispersion needs to be zero. The only orbits $r_+=1$ which also have zero tangential velocity are purely radial orbits. These radial orbits also contribute to the mass density and the radial velocity dispersion at radii $r<1$. The requirement of reproducing the density at large radii using only radial orbits without creating a mass excess at small radii is impossible. The only possible way is to add orbits with negative weight to the orbital mixture. If we require that, on top of the density profile, we also need to ensure that the velocity distribution is isotropic at all radii, we end up with the curious DF given by Eq.~(\ref{df-uniformdensity}). 

Apart from the fact that the DF is negative over a large part of phase space and thus non-physical, it is also remarkable that the Eddington formula automatically generates an energy truncation. While we explicitly used a truncation in radius rather than a truncation in binding energy as starting point, we effectively still end up with the latter if we require velocity anisotropy. A truncation in radius always leads to a factor $\Theta(\Psi-\calET)$ in the augmented density. Any augmented distribution with such a factor will result in a DF that consists of a linear combination of terms that either contain a factor $\Theta(\calE-\calET)$ or a Dirac delta function $\delta(\calE-\calET)$. Any ergodic model with a truncation in radius is thus automatically truncated in binding energy.

\subsection{Radial and circular orbit models}
\label{radcirc.sec}

Based on the discussion in the previous subsection it is clear that we need to consider anisotropic DFs of the form $f(\calE,L)$ if we want to generate self-consistent dynamical models for the uniform density sphere. Unfortunately, the general inversion technique to construct general anisotropic dynamical models based on a given density profile and anisotropy profile is rather complex \citep{1962MNRAS.123..447L, 1986PhR...133..217D, 2021isd..book.....C}. The easiest way forward is to consider special assumptions on the anisotropy profile or the functional form of the DF for which the inversion simplifies.

There are two extreme cases of orbital configurations for which the DF can be calculated in a similar way as in the isotropic case. The first case is the radial orbit model, that is, a model in which only the radial orbits are populated. For a given density profile, the DF corresponding to this model can be calculated using an Eddington-like formula  \citep{1984ApJ...286...27R, 2004MNRAS.351...18B, 2016MNRAS.462..298O},
\begin{equation}
f(\calE,L) = 
\frac{\delta(L^2)}{\sqrt2\pi^2}\,\frac{\txd}{\txd\calE} \int_0^{\!\calE} \frac{h(\Psi)\,\txd\Psi}{\sqrt{\calE-\Psi}},
\end{equation}
where $h(r) = r^2 \rho(r)$. For the uniform density sphere, we find
\begin{equation}
f(\calE,L) = \frac{3\sqrt2}{8\pi^3}\,\delta(L^2)\,\frac{5-4\calE}{\sqrt{\calE-1}}\,\Theta(\calE-1)
\label{dfrad}
\end{equation}
This DF is negative for all $\tfrac54<\calE\leqslant\tfrac32$, which shows that the uniform density sphere cannot be populated purely with radial orbits. This is no surprise since only models with a density profile at least as steep as $r^{-2}$ can be supported by radial orbits only \citep{1984ApJ...286...27R, 2006ApJ...642..752A}.

At the other side of the orbital structure spectrum is the circular orbit model, in which only circular orbits are populated. For any spherical density profile, the circular orbit model is always guaranteed to yield a formally positive distribution function, which is most conveniently written in terms of the regular coordinates instead of the integrals of motion \citep{1984ApJ...286...27R, 2008gady.book.....B},
\begin{equation}
f(r,v_r,v_\txt) = \frac{\rho(r)}{2\pi\,v_{\text{circ}}(r)}\,\delta(v_r)\,\delta\Bigl(v_\txt-v_{\text{circ}}(r)\Bigr),
\end{equation}
with $v_{\text{circ}}(r) = -r\,\frac{\txd\Psi}{\txd r}(r)$ the circular velocity at radius $r$. For the uniform density sphere we find 
\begin{equation}
f(r,v_r,v_\txt) = \frac{3}{8\pi^2}\,\frac{1}{r}\,\delta(v_r)\,\delta(v_\txt-r)\,\Theta(1-r),
\end{equation}
or if we explicitly move to the integrals of motion \citep{1971Afz.....7..223B},
\begin{equation}
f(\calE,L) = \frac{3}{8\pi^3}\,\delta(3-2\calE-2L)\,\frac{\Theta(1-L)}{\sqrt{L}}.
\label{dfcirc}
\end{equation}
This DF is formally positive over the entire phase space, so it is a represents a physically viable dynamical model for the uniform density sphere. However, it remains an extreme and degenerate probability function that only populates a marginal fraction of the accessible orbit space.

\subsection{Constant anisotropy models}
\label{cani.sec}

We now extend our search to anisotropic models with a more general and less extreme orbital structure. An interesting subclass of the general family of anisotropic models are those with a DF with a power-law angular momentum dependence,
\begin{equation}
f(\calE,L) = f_\txA(\calE)\,L^{-2\beta}.
\end{equation}
These models are characterised by an anisotropy parameter $\beta(r) = \beta$ that is independent of radius. In principle, $\beta$ can take any value smaller than or equal to one, with $\beta=1$ corresponding to only radial orbits, $\beta=0$ to isotropy and $\beta\to-\infty$ to a model consisting only of circular orbits. 

For a given density profile and any value of $\beta$ not equal to a half-integer number, the DF of the constant anisotropy model can be found through a formula similar to the Eddington relation \citep{1986PhR...133..217D, 1991MNRAS.253..414C, 2006PhRvD..73b3524E, 2021isd..book.....C},
\begin{equation}
f(\calE,L) = \frac{(2\pi)^{-3/2}\,2^\beta\,L^{-2\beta}}{\Gamma(1-\lambda)\,\Gamma(1-\beta)}\,
\frac{\txd}{\txd\calE} \int_0^\calE \frac{\txd^m h}{\txd\Psi^m}\,\frac{\txd\Psi}{(\calE-\Psi)^\lambda},
\label{dfahi}
\end{equation}
where
\begin{gather}
m = \floor\left(\tfrac32-\beta\right),\\
\lambda = \tfrac32-\beta-m, \\
h(r) = r^{2\beta}\rho(r).
\label{hahi}
\end{gather}
If $\beta$ is a half-integer number, the inversion is particularly interesting: instead of an integration, we only need $m$ differentiations,
\begin{equation}
f(\calE,L) = \frac{1}{2\pi^2}\, \frac{L^{-2\beta}}{(-2\beta)!!}\,\left.\frac{\txd^m h}{\txd\Psi^m}\,\right|_{\Psi=\calE}.
\end{equation}
For the uniform density sphere, we find from expression~(\ref{UDS-Psi})
\begin{equation}
r(\Psi) = \begin{cases}
\;\Psi^{-1} & \quad\text{for }0<\Psi\leqslant1,
\\
\;\sqrt{3-2\Psi} & \quad\text{for }1\leqslant\Psi\leqslant\frac32.
\end{cases}
\end{equation}
and thus
\begin{equation}
h(\Psi) = \frac{3\,(3-2\Psi)^\beta}{4\pi}\,\Theta(\Psi-1).
\end{equation}
Using this expression for $h$, we end up with DFs with similar characteristics as the ergodic DF. For any value of $\beta\leqslant0$, the DF is a linear combination of terms that contain as a factor either a step function $\Theta(\calE-1)$, a Dirac delta function $\delta(\calE-1)$, or a higher-order distributional derivative of the Dirac delta function $\delta^{(n)}(\calE-1)$. For the same reasons as the isotropic case, any constant-anisotropy model truncated in radius is automatically truncated in binding energy. Moreover, it turns out that none of these constant-anisotropy DFs is positive over the entire phase space. There is hence no constant anisotropy model that can self-consistently generate a sphere of uniform density.

We can illustrate this with a number of explicit, but only formal, expressions for specific values of $\beta$. The simplest cases from a mathematical point of view are those corresponding to $\beta=\tfrac12$ and $\beta=-\tfrac12$. The former case corresponds to a mildly anisotropic dynamical model and the DF is found almost trivially,
\begin{equation}
f(\calE,L) = \frac{3}{8\pi^3}\,\frac{1}{L} \left[\delta(\calE-1)-\frac{\Theta(\calE-1)}{\sqrt{3-2\calE}}\right],
\label{dfca05}
\end{equation}
This DF is similar to the ergodic model in the sense that it shows an infinite peak at $\calE=0$ and negative values for all binding energies $1<\calE\leqslant\tfrac32$. 

The case $\beta=-\tfrac12$ corresponds to a mildly tangential model. The DF is found using two derivatives of the augmented density,
\begin{equation}
f(\calE,L) = \frac{3L}{8\pi^3}\left[\frac{\delta'(\calE-1)}{\sqrt{3-2\calE}} + \frac{2\,\delta(\calE-1)}{(3-2\calE)^{3/2}} + \frac{3\,\Theta(\calE-1)}{(3-2\calE)^{5/2}}\right].
\label{dfcam05}
\end{equation}
In this formula, $\delta'(x)$ represents the doublet function, that is, the distributional derivative of the Dirac delta function \citep[e.g.,][]{9418536}. The DF~(\ref{dfcam05}) is even more curious than the DFs (\ref{df-uniformdensity}) and (\ref{dfca05}) corresponding to the isotropic and radially anisotropic models, respectively. In any case, none of these DFs represent a realistic probability density in phase space. The correctness of the expressions (\ref{dfca05}) and (\ref{dfcam05}) can be checked by integrating it over velocity space, taking into account the properties of the Dirac delta and doublet function. One indeed recovers the correct solution (\ref{UDS-rho}).

\subsection{Osipkov-Merritt models}
\label{OM.sec}

A constant anisotropy profile is not the only possibility to generate anisotropic dynamical models with an analytical DF. An interesting and popular alternative is the family of Osipkov-Merritt models. \citet{1979PAZh....5...77O} and \citet{1985AJ.....90.1027M} independently presented an inversion method for spherical models assuming a spheroidal velocity distribution. For any spherical density profile, two one-parameter families of anisotropic dynamical models can be constructed in which the distribution function can be calculated using an inversion algorithm very similar to the Eddington equation. The first family, denoted as type I by \citet{1985AJ.....90.1027M}, consists of models with an anisotropy profile that gradually changes from isotropic in the centre to fully radial at large radii. These radial Osipkov-Merritt models have been explored for many different density profiles \citep[e.g.,][]{1985MNRAS.214P..25M, 1995MNRAS.276.1131C, 1997A&A...321..724C, 2001MNRAS.321..155L, 2002A&A...393..485B}.

For the uniform density sphere, we can, however, immediately discard these models. In terms of physical consistency, radially anisotropic dynamical models are more demanding than isotropic models. Compared to circular orbits, radial orbits more easily create an excess at small radii that cannot be compensated except by adding orbits with negative weights. If a certain potential-density pair cannot be supported by an isotropic distribution, there is no hope that it can be supported by a more radially anisotropic one. Since the isotropic model for the uniform density sphere is inconsistent (Sec.~{\ref{iso.sec}}), there is no doubt that the same accounts for all radially anisotropic Osipkov-Merritt models. 

The second family of models with an ellipsoidal velocity distribution, denoted as type II by \citet{1985AJ.....90.1027M}, is characterised by an anisotropy profile that changes from isotropic in the centre to completely tangential at $r=r_\txa$, with $r_\txa$ a free parameter called the anisotropy radius. Beyond $r=r_\txa$ the model can be continued using purely circular orbits. For models with an infinite radial extent, which includes the vast majority of the analytical models, the transition at $r=r_\txa$ makes these type~II models awkward and unphysical, which explains their relative unpopularity compared to the type~I models. For density distributions with a finite radial extent, however, these problems are not present, as long as the anisotropy radius is larger than or equal to the truncation radius. The uniform density sphere is the ultimate example. Since tangentially anisotropic models have a preference for almost circular orbits which do not generate a large excess at smaller radii, they can more easily support a given potential-density pair than isotropic or radially anisotropic models. There is hence hope that we can construct physically consistent tangential Osipkov-Merritt models for the uniform density sphere. 

For type~II Osipkov-Merritt models, the distribution function depends on $\calE$ and $L$ only through the combination
\begin{equation}
Q = \calE + \frac{L^2}{2r_\txa^2}.
\label{Q-}
\end{equation}
For any spherical density profile and anisotropy radius, the distribution function can be obtained using an Eddington-like inversion relation \citep{1985AJ.....90.1027M, 2021isd..book.....C}:
\begin{equation}
f(\calE,L) = \frac{\sqrt2}{4\pi^2}\,\frac{\txd}{\txd Q} \int_0^Q \frac{\txd h}{\txd\Psi}\,
\frac{\txd\Psi}{\sqrt{Q-\Psi}},
\label{om}
\end{equation}
with
\begin{equation}
h(r) = \left(1-\frac{r^2}{r_\txa^2}\right)\rho(r).
\label{hom}
\end{equation}
For the uniform density sphere, we find for $r_\txa\geqslant1$,
\begin{multline}
f(\calE,L) = \frac{3\sqrt2}{16\pi^3}
\left[\left(1-\frac{1}{r_\txa^2}\right) \frac{\delta(Q-1)}{\sqrt{Q-1}}
\right.
\\-
\left.
\left(1+\frac{3-4Q}{r_\txa^2}\right) \frac{\Theta(Q-1)}{2\,(Q-1)^{3/2}}\right].
\label{dfQ}
\end{multline}
This distribution has a strong resemblance to the ergodic DF (\ref{df-uniformdensity}), with the main difference that $\calE$ is replaced by the quantity $Q$. Actually, in the limit $r_{\text{a}}\to\infty$, we see from Eq.~(\ref{Q-}) that $Q\to\calE$ and it is easy to verify that the DF (\ref{dfQ}) reduces to the expression~(\ref{df-uniformdensity}). For any value $r_{\text{a}}>1$, formula~(\ref{dfQ}) does not represent a physically acceptable probability function: it is always negative for $Q\gtrsim1$. The bottomline is that the uniform density sphere cannot be supported by any Osipkov-Merritt model with $r_\txa>1$.

\begin{figure*}
\includegraphics[width=\textwidth]{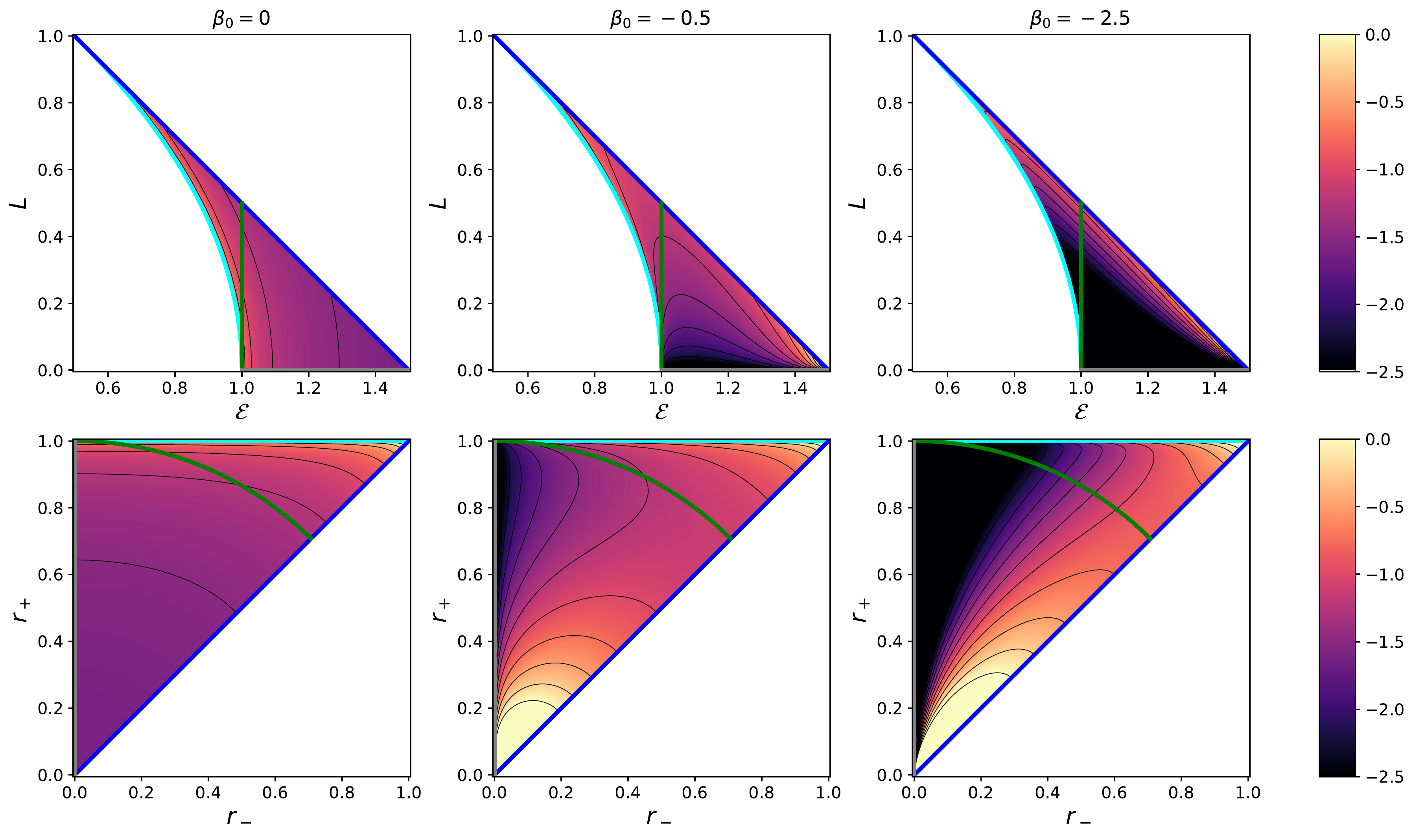}%
\caption{Distribution functions for our family of self-consistent dynamical models for the uniform density sphere in the $(\calE,L)$ parameter space (top row) and in the turning point space (bottom row). The different columns correspond to different values for the central anisotropy, ranging from $\beta_0=0$ (tangential Osipkov-Merritt model, left) to $\beta_0=-2.5$ (right). The thick coloured solid lines are the same as in Fig.~{\ref{paramspace.fig}}.}
\label{dfs.fig}
\end{figure*}

The exception is the limiting case with $r_{\text{a}}=1$. If we set $r_{\text{a}}=1$ in Eq.~(\ref{dfQ}), the first term between the square brackets disappears, and we obtain a model with a surprisingly simple DF, 
\begin{equation}
f(\calE,L) = \frac{3\sqrt2}{8\pi^3} \frac{\Theta(Q-1)}{\sqrt{Q-1}}.
\label{dfOM}
\end{equation}
This DF is positive for all values of $\calE$ and $L$, which implies that this model represents a physically acceptable dynamical model for the uniform density sphere \citep{1974SvA....17..460P, 1979PAZh....5...77O}. In the left column of Fig.~{\ref{dfs.fig}} we show this DF in the $(\calE,L)$ parameter space (top panel) and in turning point space (bottom panel). Apart from being physically consistent, this dynamical model has the interesting characteristic that it is not truncated in binding energy.  On the contrary, the DF is nonzero over the entire allowed region in phase space. In other words: all bound orbits with apocentre $r_+\leqslant1$ are populated. 

Integrating the DF over velocity space we can easily recover the density profile \citep{2021Ap.....64..219B}. When we first multiply the DF with either $v_r^2$ or $v_\txt^2$, and then integrate over velocity space, we find the radial and tangential velocity dispersion profiles,
\begin{gather}
\sigma_r^2(r) = \frac{1-r^2}{4},
\label{OMsigmar2} \\
\sigma_\txt^2(r) = \frac12.
\label{OMsigmat2}
\end{gather}
The same expressions can also be found from the solution of the Jeans equations \citep{2005MNRAS.362...95M, 2021A&A...652A..36B}. Interestingly, the tangential velocity dispersion is constant as a function of radius, whereas the radial dispersion decreases from a finite value in the centre to zero at the truncation radius. The anisotropy profile is
\begin{equation}
\beta(r) \equiv 1-\frac{\sigma_\txt^2(r)}{2\sigma_r^2(r)} = -\frac{r^2}{1-r^2}.
\end{equation}
This expression is in agreement with the general formula applicable to type~II Osipkov-Merritt models: the models are isotropic in the centre and fully tangential at the outer radius.

\section{A family of self-consistent dynamical models}
\label{newfam.sec}

\subsection{Model construction and DFs}

In the previous section we have applied the most common inversion techniques to investigate potential dynamical models for the uniform density sphere. Most orbital anisotropy profiles lead to nonphysical DFs that are negative in at some part of phase space. We have found two models with a positive DF that can self-consistently support the uniform density sphere. In Sec.~{\ref{radcirc.sec}} we discussed the circular orbit model that is guaranteed to yield a positive DF for every spherical density profile. In Sec.~{\ref{OM.sec}} we have discussed the tangential type~II Osipikov-Merritt model with $r_\txa=1$, characterised by an anisotropy profile that changes from isotropic in the centre to fully tangential at the outer radius.

The existence of these two models suggests that it might be possible to construct a family of physically viable dynamical models for the uniform density sphere in which the anisotropy gradually changes from an arbitrary value $\beta_0$ at the centre to fully tangential at the truncation radius. The case $\beta_0=0$ would then correspond to the model discussed in Sec.~{\ref{OM.sec}}, and the case $\beta_0\to-\infty$ to the circular orbit model.

To construct such a family of models we take inspiration from the work of \citet{1991MNRAS.253..414C}. He generalised the type~I Osipkov-Merritt models to dynamical models in which the anisotropy can assume any value $\beta_0$ in the centre and becomes completely radial at large radii. The Osipkov-Merritt type~I models are a special case corresponding to $\beta_0=0$. He presented families of models for the Plummer and Jaffe density profiles, and other Cuddeford models have been presented and analysed for other common density profiles \citep{2002A&A...393..485B, 2010MNRAS.401.1091C}.

\citet{1991MNRAS.253..414C} limited his analysis to an extension of the type~I Osipkov-Merritt systems. The reason is the same as indicated in the previous section: type~II Osipkov-Merritt models all contain a physically unrealistic discontinuity at $r=r_\txa$. This is a perfectly valid argument for models with an infinite extent, but does not apply to models with a radial truncation such as the uniform density sphere. It turns out that it is perfectly possible to generalise the methodology and the equations from \citet{1991MNRAS.253..414C} to type~II systems. Concretely, given a density profile $\rho(r)$ truncated at $r=r_\txT$, an anisotropy radius $r_\txa\geqslant r_\txT$, and a central anisotropy $\beta_0$, we can define a dynamical model consistent with the density profile $\rho(r)$, and with anisotropy profile
\begin{equation}
\beta(r) = \frac{r^2-\beta_0\,r_\txa^2}{r^2-r_\txa^2}.
\end{equation}
The DF of this model can be calculated as
\begin{equation}
f(\calE,L) 
= 
\frac{(2\pi)^{-3/2}\,2^{\beta_0}\,L^{-2\beta_0}}{\Gamma(1-\lambda)\,\Gamma(1-\beta_0)}\,
\frac{\txd}{\txd Q}
\int_0^Q \frac{\txd^m h}{\txd\Psi^m}\,\frac{\txd Q}{(Q-\Psi)^\lambda},
\label{cudd}
\end{equation}
where $Q$ is given by expression~(\ref{Q-}), and
\begin{gather}
m = \floor\left(\tfrac32-\beta_0\right),\\
\lambda = \tfrac32-\beta_0-m,\\
h(\Psi) = r^{2\beta_0}\left(1-\frac{r^2}{r_\txa^2}\right)^{1-\beta_0} \rho(r).
\label{hcudd}
\end{gather}
This expression is only valid when $\beta_0$ is not a half-integer number. If $\beta_0$ is a half-integer number, we have the simpler inversion formula
\begin{equation}
f(\calE,L) = \frac{1}{2\pi^2}\, \frac{L^{-2\beta_0}}{(-2\beta_0)!!}\,\left.\frac{\txd^m h}{\txd\Psi^m}\,\right|_{\Psi=Q}
\end{equation}
instead. It is clear that these formulae are very similar to the formulae for the DF of the constant anisotropy models discussed in Sec.~{\ref{cani.sec}}. Actually in the limit $r_\txa\to\infty$, we have already noted that $Q\to\calE$ and we recover the formulae (\ref{dfahi})--(\ref{hahi}) if we set $\beta=\beta_0$. On the other hand, for $\beta_0=0$, the formulae (\ref{cudd})--(\ref{hcudd}) reduce to the Osipkov-Merritt inversion formulae~(\ref{om})--(\ref{hom}).

In principle, this framework allows us to consider a two-parameter family of dynamical models for the uniform density sphere, with free parameters $\beta_0\leqslant1$ and $r_\txa\geqslant1$. However, based on our analysis of the tangential Osikpov-Merritt models, it can expected that models with $r_\txa>1$ will not result in DFs that are positive over the entire phase space. It turns out that, indeed, all models with $r_\txa>1$ are unphysical as they are negative in some region of phase space. If we restrict ourselves to $r_\txa=1$, we have a one-parameter family of models with anisotropy profile
\begin{equation}
\beta(r) = \frac{\beta_0-r^2}{1-r^2}.
\label{aniprof}
\end{equation}
The anisotropy is equal to $\beta_0$ in the centre and it gradually decreases for increasing radius, until the outer region of the sphere are completely dominated by nearly circular orbits. 

For any non-half-integer value of $\beta_0$, expression~(\ref{cudd}) leads to a rather complex expression of the DF that involves rational and hypergeometric functions. For integer values of $\beta_0$, this expression can be written in terms of elementary functions; for quarter-integer values, we obtain a combination of elementary functions and complete elliptic integrals. These expressions quickly become quite cumbersome as $m$ increases, however. The most interesting and simple cases are those where $\beta_0$ is a half-integer number, as the calculation of the DF then only involves multiple differentiations. The simplest cases are $\beta_0=\tfrac12$, for which one easily finds
\begin{equation}
f(\calE,L) = \frac{3}{8\sqrt2\,\pi^3}\,\frac{1}{L}\,\frac{5-4Q}{\sqrt{3-2Q}}\,\frac{\Theta(Q-1)}{\sqrt{Q-1}},
\label{df-c05}
\end{equation}
and $\beta_0=-\tfrac12$, for which
\begin{equation}
f(\calE,L) = \frac{9}{8\sqrt2\,\pi^3}\,\frac{L}{(3-2Q)^{5/2}}\,\frac{\Theta(Q-1)}{\sqrt{Q-1}}.
\end{equation}
Every time we decrease $\beta_0$ by one unit, the expression of the DF becomes slightly more complex. For example, for $\beta_0=-\tfrac32$ we have
\begin{equation}
f(\calE,L) = \frac{15}{8\sqrt2\,\pi^3}\,\frac{L^3\,(12Q-11)}{(3-2Q)^{9/2}}\,\frac{\Theta(Q-1)}{\sqrt{Q-1}},
\end{equation}
and for $\beta_0=-\tfrac52$,
\begin{equation}
f(\calE,L) = \frac{21}{8\sqrt2\,\pi^3}\,\frac{L^5\,(64Q^2-96Q+33)}{(3-2Q)^{13/2}}\,\frac{\Theta(Q-1)}{\sqrt{Q-1}}.
\end{equation}

The crucial issue is the physical consistency, or in other words: for which values of $\beta_0$ is the DF positive of the entire phase space? It turns out that all models with a radial central anisotropy, that is with $\beta_0>0$, are nonphysical, as their DF becomes negative for the highest values of $Q$. The $\beta_0=\tfrac12$ model with expression~(\ref{df-c05}) as DF illustrates this: the DF is negative for all $\tfrac54<Q<\tfrac32$. For all models with $\beta_0\leqslant0$, however, the DF is positive over the entire phase space. We hence have a family of models with varying central anisotropy in which each model self-consistently generates the uniform density sphere. Moreover, for each model the DF is nonzero over the entire accessible phase space, implying that all possible bound orbits are populated. 

In the central and right columns of Fig.~{\ref{dfs.fig}} we compare the DFs corresponding to $\beta_0=0$, $\beta_0=-\tfrac12$ and $\beta_0=-\tfrac52$. In the outskirts, that is, for apocentres $r_+\lesssim1$, the three DFs have a similar behaviour. Indeed, irrespective of the value of $\beta_0$, the outer region is dominated by tangential orbits. The three models do differ in the central regions, with an increasing contribution of tangential orbits as $\beta_0$ grows more negative. In the limit $\beta_0\to-\infty$, the model reduces to the degenerate model in which all orbits are purely circular orbits. In this case, the DF is identically zero over the entire parameter space, except along the blue line in Fig.~{\ref{dfs.fig}}. 

\subsection{Velocity dispersions and the virial theorem}

Instead of focusing on the DF, it is often more intuitive to study the moments of the DF with respect to the velocities, with the radial and tangential velocity dispersion profiles as the most important cases for a spherical dynamical model. For a general anisotropic spherical dynamical model, they can be found as
\begin{gather}
\sigma_r^2(r) = \frac{1}{\rho(r)} \int f(\calE,L)\,v_r^2\,\txd\bfv,\\
\sigma_\txt^2(r) = \frac{1}{\rho(r)} \int f(\calE,L)\,v_\txt^2\,\txd\bfv.
\end{gather}
For a given anisotropy profile, the velocity dispersion profiles can also be determined from the Jeans equations \citep[for a detailed discussion, see][]{2021isd..book.....C}. For the specific case of the tangential Cuddeford-like models with anisotropy profile~(\ref{aniprof}), the appropriate formula can be shown to be
\begin{equation}
\sigma_r^2(r) 
= 
\frac{r^{-2\beta_0}}{(1-r^2)^{1-\beta_0}}\,\frac{1}{\rho(r)}
\int_r^1 \rho(u)\,M(u) \left(\frac{1-u^2}{u^2}\right)^{1-\beta_0}\, \txd u.
\end{equation}
Using expressions~(\ref{UDS-rho}) and (\ref{UDS-M}), we find the general expression
\begin{multline}
\sigma_r^2(r) 
=
\frac{1}{2}\,r^{-2\beta_0} \left(1-r^2\right)^{\beta_0-1}
\\
\times\Bigl[\txB(1+\beta_0,2-\beta_0)-\txB_{r^2}(1+\beta_0,2-\beta_0)\Bigr],
\label{raddisp}
\end{multline}
where $\txB(a,b)$ and $\txB_x(a,b)$ represent the complete and the incomplete beta functions, respectively. The corresponding tangential distribution can be calculated via
\begin{equation}
\sigma_\txt^2(r) 
= 
2\,\bigl[1-\beta(r)\bigr]\,\sigma_r^2(r) 
=
\frac{2\,(1-\beta_0)}{1-r^2}\,\sigma_r^2(r).
\label{tandisp}
\end{equation}
For all integer and half-integer values of $\beta_0$, the dispersion profiles can be expressed in terms of elementary functions. For the limiting case $\beta_0=0$, we easily recover the expressions (\ref{OMsigmar2}) and (\ref{OMsigmat2}). 

\begin{figure*}
\includegraphics[width=0.9\textwidth]{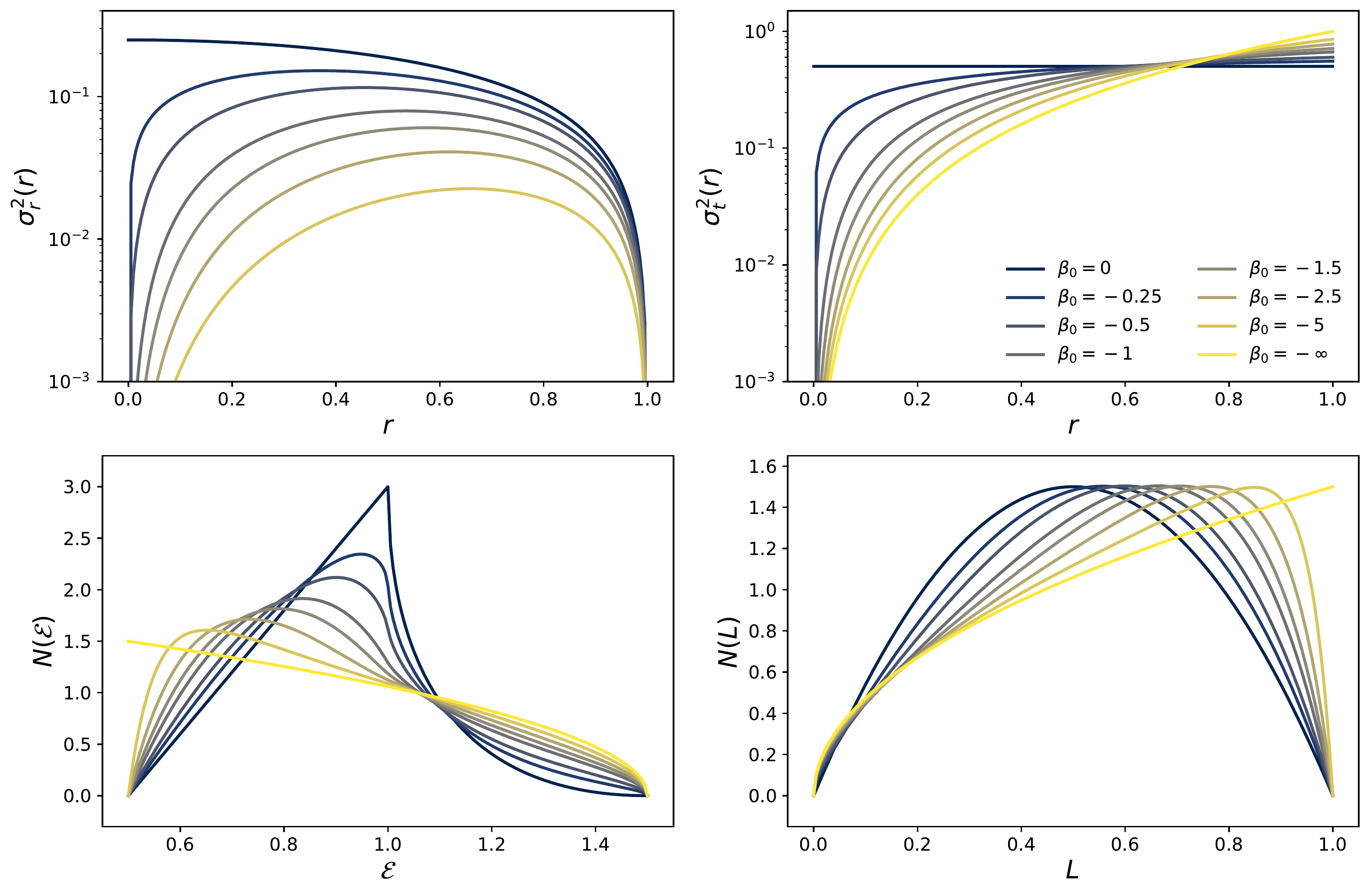}\hspace*{2em}%
\caption{Dynamical properties for our family of self-consistent dynamical models for the uniform density sphere: radial velocity dispersion profile (upper left), tangential velocity dispersion profile (upper right), differential energy distribution (lower left), and differential angular momentum distribution (lower right). Different colours correspond to different values for the central anisotropy, ranging from $\beta_0=0$ (the tangential Osipkov-Merritt model) to $\beta_0\to-\infty$ (the circular orbit model).}
\label{dynamics.fig}
\end{figure*}

The upper left panel of Fig.~{\ref{dynamics.fig}} shows the radial velocity dispersion profiles for a selection of models with different values of the central anisotropy $\beta_0$. For any fixed value of $\beta_0$, the radial dispersion profile starts at zero in the centre, subsequently increases to a maximum value, and ultimately decreases again to reach zero at the outer edge of the sphere. The only exception is the limiting case $\beta_0=0$ with an isotropic core, for which the radial dispersion at the centre converges to a finite value. The asymptotic behaviour at small radii reads
\begin{equation}
\sigma_r^2(r) \sim
\begin{cases}
\;\dfrac14\,\Gamma(1+\beta_0)\,\Gamma(2-\beta_0)\,r^{-2\beta_0} & \quad{\text{for }}-1<\beta_0\leqslant 0, \\[1em]
\;\left(\ln\dfrac{1}{r}-\dfrac34\right)r^2 & \quad{\text{for }}\beta_0=-1, \\[1em]
\;\dfrac{1}{2\,(-1-\beta_0)}\,r^2 & \quad{\text{for }}\beta_0<-1.
\end{cases}
\end{equation}
Near the outer boundary of the sphere, the radial velocity dispersion linearly drops to zero for all models,
\begin{equation}
\sigma_r^2(r) \sim \frac{1-r}{2-\beta_0}.
\end{equation}
The tangential velocity dispersion profiles, shown in the upper right panel Fig.~{\ref{dynamics.fig}}, have a distinctly different behaviour. For any fixed value of $\beta_0$, the tangential dispersion increases as a function of radius, from zero at the centre to 
the maximum value
\begin{equation}
\sigma_\txt^2(1) = \frac{1-\beta_0}{2-\beta_0}.
\end{equation}
The limiting case $\beta_0=0$ model is the exception on the rule with a constant value. 

As $\beta_0$ gradually grows more negative, the model becomes gradually more tangentially anisotropic, with a gradual larger contribution of circular-like orbits. As a result, the radial (tangential) velocity dispersion systematically decreases (increases) at any radius. In the limit $\beta_0\to-\infty$, corresponding to the circular orbit model, the radial velocity dispersion is obviously zero at all radii, whereas the tangential velocity dispersion is equal to the square of the circular velocity profile,
\begin{equation}
\sigma_\txt^2(r) = v_{\text{circ}}^2(r) = r^2,
\end{equation}
a well-known result for the uniform density sphere.  

With the explicit expression for velocity dispersion profiles, we can calculate the total kinetic energy,
\begin{equation}
T_{\text{tot}} = 2\pi\int_0^\infty \rho(r)\,\left[\sigma_r^2(r) + \sigma_\txt^2\right] r^2\,{\text{d}}r.
\label{Ttotdef}
\end{equation}
After substitution of expressions~(\ref{UDS-rho}), (\ref{raddisp}) and (\ref{tandisp}) into this equation and some calculation, one obtains a simple results that does not depend on the central anisotropy $\beta_0$,
\begin{equation}
T_{\text{tot}} = \frac{3}{10}.
\label{Ttot}
\end{equation}
Comparing this expression to equation~(\ref{Wtot}) we see that the virial theorem is satisfied,
\begin{equation}
W_{\text{tot}} + 2\,T_{\text{tot}} = 0,
\end{equation}
as expected.

\subsection{Differential energy and differential angular momentum distributions}

While the DF contains all possible dynamical information, it is important to realise that $f(\calE,L)$ does not represent the probability distribution in $(\calE,L)$ space, that is, the distribution of mass per unit binding energy and unit angular momentum. This quantity, denoted as
\begin{equation}
N(\calE,L) \equiv \frac{\txd^2 M}{\txd\calE\,\txd L}(\calE,L),
\end{equation}
is a very relevant diagnostic of a spherical stellar system too, and becomes particularly interesting if one regards a stellar system as a superposition of orbits, each of them characterised by a specific value of $\calE$ and $L$. 

To calculate $N(\calE,L)$ for an anisotropic dynamical model, we need to multiply the DF $f(\calE,L)$ by the density-of-states function $g(\calE,L)$, which represents the phase-space volume per unit binding energy and per unit angular momentum. The latter quantity is calculated as
\begin{equation}
g(\calE,L) = 16\pi^2\,L \int_{r_-}^{r_+} \frac{r\,\txd r}{\sqrt{2r^2 [\Psi(r)-\calE]-L^2}},
\label{gEL}
\end{equation}
where $r_-$ and $r_+$ are the pericentre and apocentre of an orbit with binding energy $\calE$ and angular momentum $L$ \citep[e.g.,][]{1986PhR...133..217D, 2009ApJ...690.1280V}. For the uniform density sphere, we find a very simple expression for $g(\calE,L)$, at least for orbits corresponding to $r_+\leqslant1$,
\begin{equation}
g(\calE,L) = 8\pi^3\,L.
\label{gEL}
\end{equation}

Rather than the joint distribution $N(\calE,L)$, the individual distributions of $N(\calE)$ and $N(L)$ are more often investigated. The differential energy distribution (DED) $N(\calE)$ can be considered as the most fundamental partitioning of a stellar system \citep{1982MNRAS.200..951B, 2007LNP...729..297E, 2010ApJ...722..851H}. It is calculated by marginalising $N(\calE,L)$ over angular momentum,
\begin{equation}
N(\calE) \equiv \frac{\txd M}{\txd\calE}(\calE) = \int N(\calE,L)\,\txd L.
\label{NEdef}
\end{equation}
The DED corresponding to different values of $\beta_0$ is shown in the lower left panel of Fig.~{\ref{dynamics.fig}}. In all cases, the DED is nonzero for binding energies in the interval $\tfrac12\leqslant\calE\leqslant\tfrac32$, showing again that all possible binding energies allowed are populated. It varies smoothly and systematically as $\beta_0$ changes. The limiting values can be calculated analytically. For the tangential Osipkov-Merritt model with $\beta_0=0$, the DF is given by expression~(\ref{dfOM}), and if we combine this expression with (\ref{gEL}) and (\ref{NEdef}), we find
\begin{equation}
N(\calE) = 3\left(2\calE-1\right) -6\sqrt2\,\sqrt{\calE-1}\,\Theta(\calE-1).
\end{equation}
For the circular orbit model with DF~(\ref{dfcirc}) we find 
\begin{equation}
N(\calE) = \frac{3\sqrt2}{4}\,\sqrt{3-2\calE}. \\
\end{equation}
Looking in more detail at the DEDs shown in the lower left panel of Fig.~{\ref{dynamics.fig}}, one can see that the mean value of the DED of each model is exactly the same: the mean binding energy is equal to
\begin{equation}
\langle \calE \rangle \equiv \frac{1}{M} \int N(\calE)\,\calE\,\txd\calE = \frac{9}{10},
\label{meancalE}
\end{equation}
irrespective of the value of $\beta_0$. In fact, even for the nonphysical models with DFs that are negative in some part of phase space, such as expressions~(\ref{df-uniformdensity}), (\ref{dfrad}), (\ref{dfca05}), (\ref{dfcam05}) or (\ref{dfQ}), we find exactly the same value if we formally calculate the DED and subsequently its mean value $\langle\calE\rangle$. This is in agreement with \citet{2021A&A...653A.140B}, who demonstrated that the mean binding energy of any steady-state dynamical model is independent of the orbital structure. If we define the total integrated binding energy as 
\begin{equation}
B_{\text{tot}} = M\langle\calE\rangle,
\label{Btot}
\end{equation}
and we compare the expressions (\ref{meancalE}), (\ref{Btot}), (\ref{Wtot}), and (\ref{Ttot}), we find
\begin{equation}
B_{\text{tot}} = 3\,T_{\text{tot}} = -\frac32\,W_{\text{tot}} = -3\,E_{\text{tot}}.
\label{energyrelation}
\end{equation}
We recover the general conclusion by \citet{2021A&A...653A.140B} that the total integrated binding energy supplements the well-known trio consisting of total kinetic energy, total potential energy, and total energy on an equal footing. Knowledge of any one out of these four energies suffices to calculate the other three.

Interestingly, for the uniform density sphere, the mean binding energy is smaller than the truncation energy. This immediately shows that, for this particular case, no physical models with an energy truncation are possible: it is impossible for a probability function $N(\calE)$ defined on an interval $1\leqslant\calE\leqslant\tfrac32$ to have a mean value below the lower limit of the interval. So not only the ergodic, the radial orbit, and the constant anisotropy models discussed in Secs.~{\ref{iso.sec}}, {\ref{radcirc.sec}}, and~{\ref{cani.sec}} are nonphysical, as we have shown, but the same accounts for any dynamical model with a binding energy truncation at $\calE=1$.

Another DED-related finding of \citet{2021A&A...653A.140B} was that, for the Hernquist model, the width of the DED varies systematically with the anisotropy: radially anisotropic models tend to prefer more average binding energies, whereas models with a more tangential orbital distribution favour a broader distribution with more extreme binding energies. The same behaviour is noted here for the uniform density sphere: the shape of the DED gradually changes from more peaked for the Osipkov-Merritt model to broader for the circular orbit model. This can be quantified by means of the standard deviation of the DED. For our set of models, this standard deviation gradually increases from 0.117 for $\beta_0=0$ to 0.641 for $\beta_0\to-\infty$.

In a similar way as one defines the DED, one can consider the differential angular momentum distribution (DAMD) $N(L)$. This quantity has primarily been investigated in the frame of the $N$-body simulations of dark matter haloes \citep[e.g.,][]{2001ApJ...555..240B, 2005ApJ...628...21S, 2012ApJ...750..107S}. It is calculated by marginalising $N(\calE,L)$ over binding energy,
\begin{equation}
N(L) \equiv \frac{\txd M}{\txd L}(L) = \int N(\calE,L)\,\txd\calE.
\end{equation}
The lower right panel of Fig.~{\ref{dynamics.fig}} shows the DAMD for our family of uniform density sphere models for different vales of the central anisotropy $\beta_0$. For every model, the DAMD is defined in the interval $0\leqslant L \leqslant 1$, and it varies smoothly as a function of $\beta_0$. For the models at the extremities of the allowed range in $\beta_0$ we can calculate the DAMD explicitly: for the Osipkov-Merritt model $\beta_0=0$ we have 
\begin{equation}
N(L) = 6L\,(L-1),
\end{equation}
and for the circular orbit model 
\begin{equation}
N(L) = \frac32\,\sqrt{L}.
\end{equation}
Contrary to the DEDs, the DAMDs do not all share the same mean value: the mean angular momentum of the models changes systematically to higher values as the central anisotropy changes from isotropic to completely tangential. For the $\beta_0=0$ model we have $\langle L \rangle = \frac12$, for the circular orbit model, $\langle L \rangle = \frac35$.

\section{Discussion and conclusion}
\label{sumdisc.sec}

As discussed in the Introduction section, they standard way to generate dynamical models with a finite extent is the application of an energy truncation to a distribution function $f(\calE,L)$. This approach has the disadvantage, however, that it does not automatically (and generally not at all) lead to models with an analytical potential or density. Moreover, energy truncated models never populate the entire range of possible orbits: they always leave a fraction of the possible parameter space empty. We have explored another route to dynamical models with a finite extent: rather than starting from a truncation in binding energy, we have started from a truncation in radius. We have focused on the simplest truncated density profile, the uniform density sphere.

In Sec.~{\ref{quest.sec}} we have explored the most common inversion techniques to generate DFs from a given density profile, covering a large range of possible anisotropy profiles. Thanks to the comprehensibility of the uniform density sphere, many of the inversion formulae lead to analytical formulae for the corresponding DFs. Unfortunately, in many cases, these DFs turned out to be formal expressions rather than physically viable probability density functions in phase space.
\begin{enumerate}
\item It was already known that the uniform density sphere cannot be supported by an isotropic velocity distribution \citep{Zeldovich72, 1979PAZh....5...77O, 2008gady.book.....B}. We have derived an explicit formal expression for the DF, which is negative over almost the entire phase space. Moreover, under the assumption of isotropy, a truncation in radius automatically translates to a truncation in binding energy. 
\item The radial orbit model is also nonphysical, as expected for a spherical model with a constant density core \citep{1984ApJ...286...27R, 2006ApJ...642..752A}. At the other extreme end of the orbital structure spectrum, the circular orbit model has a positive DF, and hence represents a physically viable dynamical model, as expected \citep{1971Afz.....7..223B, 1984ApJ...286...27R, 2008gady.book.....B}. It is an extreme model that populates only a marginal fraction of the possible orbits, however.
\item For anisotropic dynamical models with a constant anisotropy profile we find a similar situation as for the ergodic model: there is no model with constant anisotropy that can generate the uniform density sphere, and a truncation in radius automatically generates a truncation in binding energy.
\item Since the isotropic DF is nonphysical, all radial Osipkov-Merritt models are nonphysical as well. 
\item Tangentially anisotropic Osipkov-Merritt models are less popular than their radially anisotropic counterparts, because these models are characterised by a physical discontinuity at the anisotropy radius. For models with a finite extent, however, this discontinuity is irrelevant if the anisotropy radius is not smaller than the truncation radius. For the uniform density sphere, we find that all tangential Osipkov-Merritt models with $r_\txa>1$ are nonphysical, whereas the model with $r_\txa=1$ results in a physically viable model with a very simple DF that populates all possible orbits \citep{1974SvA....17..460P, 1979PAZh....5...77O, 2021Ap.....64..219B}.    
\end{enumerate}

In Sec.~{\ref{newfam.sec}} we have generated and discussed a one-parameter family of dynamical models. Each member of this family is characterised by an anisotropy profile that smoothly decreases from an arbitrary value $\beta_0\leqslant0$ at the centre to completely tangential at the truncation radius. This family of models forms a sequence between the tangential Osipkov-Merritt model and the circular orbit model. To calculate the DF of these models, we have generalised the inversion method of \citet{1991MNRAS.253..414C} to tangentially anisotropic models. The resulting formulae are very similar as in the original case corresponding to radially anisotropic models, and become particularly simple for models with half-integer values of $\beta_0$. When applied to the uniform density model, we have found that all models with $\beta_0\leqslant0$ have a positive DF over the entire phase space, and that all possible orbits are populated. We have discussed a number of dynamical properties of this family of models in detail, including the velocity dispersion profiles, the differential energy distribution, the differential angular momentum distribution and the energy budget. 

In our view, the most important result of this study is that it shows that one can generate nontrivial self-consistent dynamical models based on preset density profile with a finite extent. An energy truncation is hence not the only way forward, and also not necessarily the best way, as already advocated by \citet{1988ApJ...325..566K} more than 30 years ago. On the other hand, this study can and probably should be extended and improved in several directions. One of the obvious extensions is to investigate other density profiles. The uniform density sphere has the advantage that its density profile is so simple that many calculations can be performed analytically. It is quite an extreme model, however, even within the general family of models with a finite extent. The fact that the density does not fall with increasing radius makes it very hard to populate this model with radial orbits, as such orbits always create a density excess at smaller radii. Moreover, the uniform density sphere is not only characterised by a finite extent, but by a discontinuity in the density profile at the truncation radius. This discontinuity translates to a Heaviside step function in the augmented density, and a Dirac delta function in its derivative. It is not difficult to envision a set of models with a finite extent but without density discontinuity, and it remains to be seen how this difference translates to the consistency in the dynamical structure. We plan to investigate these issues in future work.

\section*{Acknowledgements}

We thank the anonymous referee for a swift and constructive referee report.

\section*{Data availability}

No data were used in this research. The data generated and the plotting routines will be shared on reasonable request to the corresponding author.

\bibliographystyle{mnras}
\bibliography{UDSbib}

\bsp	
\label{lastpage}
\end{document}